\documentclass[preprint,showpacs]{revtex4-1}
\usepackage{amssymb}
\usepackage{amsmath}
\usepackage{graphicx}
\usepackage[colorlinks=true, citecolor=blue, urlcolor = blue, linkcolor= red, bookmarks=true]{hyperref}
\usepackage{orcidlink}

 \def\frac#1#2{{\textstyle{{#1}\over
			{#2}}}} 
\def\lsim{\mathrel{\rlap{\lower4pt\hbox{\hskip1pt$\sim$}}
		\raise1pt\hbox{$<$}}}
\def\gsim{\mathrel{\rlap{\lower4pt\hbox{\hskip1pt$\sim$}}
		\raise1pt\hbox{$>$}}} \def\sqr#1#2{{\vcenter{\vbox{\hrule
				height.#2pt \hbox{\vrule width.#2pt height#1pt \kern#1pt \vrule
					width.#2pt} \hrule height.#2pt}}}}

\def\beq{\begin{equation}} \def\eeq{\end{equation}}
\def\beqa{\begin{eqnarray}} \def\eeqa{\end{eqnarray}}

\usepackage{hyperref}

\begin{document}

\title{ Leptogenesis in Exponential $f(R)$ Gravity Model}
\author{Suhail Khan\orcidlink{0009-0007-4941-0069}$^{1}$}
\email[]{suhail@ctp-jamia.res.in}\email[]{ suhail187148@st.jmi.ac.in}
\author{Ajay Bassi\orcidlink{0000-0001-8915-3860}$^{2}$}
\email[]{ajay@ctp-jamia.res.in}
\author{Rathin Adhikari\orcidlink{0000-0002-8764-9587}$^{1}$}
\email[]{rathin@ctp-jamia.res.in}

\affiliation{$^{1}$Centre for Theoretical Physics, Jamia Millia Islamia, New Delhi, India}
\affiliation{$^{2}$P. P. Savani University, Dhamdod, Surat, Gujarat, India}

\date{\today}
\begin{abstract}
We show that gravitational leptogenesis with dynamical $CPT$ breaking in an expanding universe can be reconciled with the exponential $f(R)$ gravity model, 
which introduces only one additional parameter $\beta$ compared to the standard $\Lambda$CDM cosmology.
This model incorporated axions as cold dark matter.  For $L$ violating interactions, we consider both a non-supersymmetric model with heavy right-handed neutrino decay and 
a supersymmetric model with sneutrino decay. For both cases, we have shown that the required baryonic asymmetry could be obtained. We have also shown the variation of decoupling temperature for lepton number violating interactions with the $\beta$ parameter in exponential  $f(R)$ gravity. Lepton number-violating model parameters are constrained with the $\beta$ through the decoupling temperature. An upper bound on the $\beta$ parameter of the exponential $f(R)$ gravity is also obtained. 

\end{abstract}
\maketitle
\section{Introduction}\label{sec:level1}
A non-vanishing baryon asymmetry is implied by the measured baryon-to-entropy ratio. This is one of the major mysteries of our Universe to this day. This  asymmetry, as obtained by the Planck collaboration, is \cite{Planck:2015fie}, 
\begin{equation}
\label{obs}
{n_b -n_{\bar{b}} \over s} \lesssim 8.6 \pm 0.1 \times 10^{-11}~,
\end{equation}
where $n_b$,\,$n_{\bar{b}}$ and $s$ are the baryon number density, anti-baryon number density, and entropy density, respectively. 
In obtaining the observed asymmetry, Sakharov's three conditions are, in general, to be satisfied: \\ 
1. Existence of baryon/lepton number-violating interactions, \\ 
2. Presence of $C$ and $CP$-violating physical processes, \\ 
3. An out-of-equilibrium condition for $B$ or $L$ violating physical processes.  

The interference term involving the amplitude of tree-level and higher-order diagrams associated with $L$ or $B$ violating processes is required to obtain the $CP$ asymmetry.

However, some alternative mechanisms to generate the baryonic or leptonic asymmetry have been suggested, in which the interference of the above-mentioned two diagrams is not required. One such example is a dynamical $CPT$-violating interaction during the expansion of the Universe~\cite{cohen1987thermodynamic,Davoudiasl:2004gf,Lambiase:2006md}. Such an interaction term creates an effective chemical potential for particles and anti-particles in thermal equilibrium, thereby producing baryonic or leptonic asymmetry through their different number density distribution functions. 

Following the work of~\cite{Davoudiasl:2004gf}, we consider the interaction between the derivative of the Ricci scalar curvature and the lepton number current $J^{\mu}$ as
\begin{equation}
\label{int}
I=\frac{1}{M_*^2}\int d^4x \sqrt{-g}\, (\partial_{\mu} R)\, J^{\mu}~,
\end{equation}
where $M_*$ is the cutoff scale of the effective theory and $R$ is the Ricci scalar curvature.
The $L$-violating interactions are required to occur in thermal equilibrium, and at a certain decoupling temperature $T_D$, the decay width $\Gamma$ of the $L$-violating process should satisfy $\Gamma \lesssim H$, where $H$ is the Hubble parameter. This condition ensures that the asymmetry generated during equilibrium survives at later stages of the cooling of the Universe. 

In our work, for the leptonic current, we will consider two different cases of lepton-number-violating decays:  
(i) heavy right-handed neutrino decay, and  
(ii) heavy right-handed sneutrino decay. 

Using Fermi-Dirac statistics in the limiting scenario where $T \gg m_f$ (with $m_f$ the fermion mass), the difference between the number densities of fermions $n_l$ and anti-fermions $\bar{n_l}$ due to a chemical potential $\mu$ is given by:

\begin{equation}
   \label{diff}
   n_l - n_{\bar l}={1 \over 6 \pi^2 } g_l T^3 \left[ \pi^2 \left( {\mu \over T} \right)  + \left(\mu \over T \right)^3 \right] 
\end{equation}
in thermal equilibrium. Here $g_l$ is the lepton intrinsic degrees of freedom, $ T$ is the temperature, $\mu$ is the chemical potential. Due to interaction in~(\ref{int}), the corresponding effective chemical potential is $\mu \sim  {\dot{R} \over M_*^2}$. The entropy density $s$, which comes from equilibrium thermodynamics, is given by 
\begin{equation}
\label{defentr}
s = {2 \pi^2 \over 45} g_*  \; T^3~,
\end{equation}
where $g_* $ is the effective number of relativistic degrees of freedom of standard model particles contributing to the entropy, and  $g_*  \approx 107$ for the evolution of the Universe. Using effective chemical potential in the leading term in Eq.~(\ref{diff}) and using entropy density in Eq.~(\ref{defentr}), the leptonic asymmetry in Eq.~(\ref{obs}) is 
\begin{equation}
\label{eta and R}
\eta_l = {n_l - n_{\bar l} \over s} \approx -{15 g_l \over 4\pi^2 g_*} {\dot{R} \over M_*^2 T}, \quad \text{at $T=T_D$}~,
\end{equation}
where $T_D$ is the temperature at which the lepton-violating interactions decouple due to the expansion of the universe.

In general relativity, for a perfect fluid, the Ricci scalar curvature is $ R \sim (1-3\omega)\rho$, where the parameters are defined as usual, like $\omega=p/ \rho$, pressure as $p$, and energy density as $\rho$.  For the radiation-dominated era, characterized by $\omega = 1/3$,  $  \partial_{t} R= \dot{R}$ vanishes, and no net baryon number asymmetry can be generated. However, in modified gravity theories, this can be evaded.
 One may consider some modification of general relativity at small and large scales for explaining late-time acceleration of the universe and inflation in the early universe, respectively. For that, in modified $f(R)$ gravity, some non-linear terms of the Ricci scalar are considered in the action.  We will consider particularly the late-time acceleration due to modified gravity. Furthermore, the role of cold dark matter as a scalar axion field $\phi(t)$ in the expansion of the universe is also considered.
 There have not been any dark matter candidates found in the laboratory as of now. However, the gravitational effects of dark matter have been widely studied, as it is essential to explain the rotation of the galaxy curve and large-scale structure formation.  One of the most promising dark matter candidates is the axion~\cite{Preskill:1982cy,Abbott:1982af,Dine:1982ah,Kim:1986ax}, a hypothetical particle put forward also to solve the strong $CP$ problem of QCD~\cite{Peccei:1977hh}.

We consider the following action
\begin{equation}
\label{fullaction}
S= \int d^4x \sqrt{-g} [\kappa^2 f(R)+\mathcal{L}]+{1 \over M_*^2}\int d^4x \sqrt{-g}(\partial_{\mu} R) J^\mu~,
\end{equation}
where $\kappa = (8\pi G)^{-1/2} = M_P/2$ and $M_P \approx 2.4\times 10^{18}$ GeV is the reduced Planck mass scale, $g$ is the metric determinant and $\mathcal{L} = \mathcal{L}_{matter}+\mathcal{L}_{dark}$ is the total Lagrangian density. $\mathcal{L}_{matter}$ is the matter Lagrangian density and $\mathcal{L}_{dark}$ is the Lagrangian density due to dark matter.

A number of viable $f(R)$ models have been 
proposed~\cite{DeFelice:2010aj,Bamba:2010iy}, including the Starobinsky model~\cite{Starobinsky:2007hu}, the Hu-Sawicki model~\cite{Hu:2007nk}, 
the Tsujikawa model~\cite{Tsujikawa:2007xu,Cen:2019lwf}, and the exponential model~\cite{Linder:2009jz,Bamba:2010ws}. These models are considered 
viable provided they satisfy several conditions~\cite{DeFelice:2010aj,Bamba:2010iy}:  (1) positivity of the effective gravitational coupling, 
(2) stability of cosmological perturbations, (3) asymptotic behavior approaching $\Lambda$CDM in the large-curvature regime, 
(4) stability of the late-time de Sitter solution, (5) consistency with the equivalence principle, and (6) compatibility with Solar-System constraints. In this work, we focus on the exponential $f(R)$ gravity model, 
which introduces only one additional parameter $\beta$ compared to the standard $\Lambda$CDM cosmology.

We explore whether it is possible to reconcile the requirement of leptonic, and hence the baryonic, asymmetry obtained through Eq.~(\ref{int}) with that of exponential $f(R)$ gravity and axion field as dark matter for the successful late-time acceleration of the Universe.
In Section~\ref{sec:levelII}, we discuss the field equations obtained after varying the action in Eq.~(\ref{fullaction}) with respect to the metric. The form of the exponential $f(R)$ term is presented and the equation of motion for the scalar axion field is also derived. Following the ansatz for $a(t)$ and using field equations in Section ~\ref{sec:levelII}, the solutions for the scalar field $\phi(t)$  and   ${\dot R(t)}$ are obtained. In Section~\ref{br}, we show the dependence of leptonic asymmetry on the $f(R)$ gravity parameter $\beta$. Finally, in Section~\ref{dis}, we present our concluding remarks.

\section{Field equations}\label{sec:levelII}
The Friedmann-Robertson-Walker (FRW) metric is considered as 
\begin{equation}
\label{FRW}
ds^{2} = dt^2 - a^2(t)d\Omega^2~,
\end{equation}
where $a(t)$ is the scale factor and $d\Omega$ is the infinitesimal volume  in co-moving coordinates.

 The exponential $f(R)$-type model has been thoroughly investigated with great success~\cite{Bamba:2010ws,Elizalde:2010ts,Odintsov:2017qif,Odintsov:2018qug} with

\begin{equation}
\label{fr}
f (R)= R-2\Lambda \left[1-e^{-\beta \, R \over 2\, \Lambda}\right]~,
\end{equation}
where $\Lambda$ is the cosmological constant and $\beta$ is a constant parameter to be determined with constraints $\Lambda>0$ and $\beta>0$ (which is automatically satisfied for the parameter ranges discussed below). This form is a smooth, one-parameter deformation of standard GR that (i) reduces to standard GR at high curvature,  and (ii) can satisfy the
standard $f(R)$ viability conditions, such as  ghost  and tachyon free for $f(R)'>0$, $f(R)''>0$.

The metric (\ref{FRW}) leads to the Ricci scalar curvature,
\begin{equation}
\label{curvature}
R = -6(\dot{H} + 2 H^2) ~,
\end{equation}
where $H \equiv \dot{a}(t)/a(t)$ is the Hubble parameter.
Considering null variation of the action~(\ref{fullaction}) with respect to the metric $g_{\mu\nu}$, the following field equations can be obtained.

The $rr$ component of the field equations reads
\begin{equation}
\label{RAY}
\left(\dot{H} + 3H^2\right)f(R)' = {{d^2f(R)'}\over{dt^2}}  + 2 H{{df(R)'}\over{dt}} - {\kappa^2 f(R) \over 2}~,
\end{equation}

Likewise,  $tt$ component is
\begin{equation}
\label{MFE}
-3(\dot{H} + H^2) f'(R) = {\rho(t) \over 2}  - 3H{{df'(R)}\over{dt}}  + {\kappa^2 f(R)\over 2}~,
\end{equation}

While the trace is
\begin{eqnarray}
\label{TR}
3{{d^2f'(R)}\over {dt^2}} + 9H {{df'(R)}\over {dt}}  - 6\left( \dot{H} + 2H^2 \right )f'(R) \nonumber
\\=2\kappa^2 f(R)-{\rho(t) \over 2}~,
\end{eqnarray}

 $\rho(t)$ is the total matter density i.e. sum of baryonic matter $\rho_m(t)$ and axion density $\rho_a(t)$ and $f'(R)$ is derivative with respect to $R$. The axion, a coherently oscillating scalar field, has been generally acknowledged as a potential candidate for the CDM from the very beginnings of dark matter research \cite{preskill1983,abbott1983,dine1983,kim1987}.
Axion is assumed to behave as a perfect fluid, with an energy-momentum tensor
\begin{equation}
T_{\mu\nu} = [p_a(t)+\rho_a(t)]u_{\mu}u_{\nu} -p_a(t)\,g_{\mu\nu}~,
\end{equation}
where $u^{\mu}$ is its four-velocity and $\rho$ and $p$ are defined as the temporal average of the oscillating scalar field (axion),
which contribute to the background fluid quantities. For the axion  scalar field \cite{Hwang:2009js}, one can write
\begin{eqnarray}
     & & \rho_a(t) = {1 \over 2} \langle \dot \phi^2 + m^2 \phi^2 \rangle, \quad
       p_a(t) = {1 \over 2} \langle \dot \phi^2 - m^2 \phi^2 \rangle,
   \label{BG-fluid} \\
   & &
       \ddot \phi + 3 H \dot \phi + m^2 \phi = 0,
   \label{BG-EOM}
 \end{eqnarray}
 where the angular bracket indicates averaging over time scale of order $m^{-1}$, where $m$ is the small mass of the axion.
 
\section{Gravitational Baryogenesis with Exponential $f(R)$ Gravity} \label{br}
To solve Eqs.~(\ref{RAY})--(\ref{BG-EOM}), the following ansatz for the scale factor 
\begin{equation}\label{alp}
 a(t) \sim \left({t \over t_0} \right) ^{\alpha}, \quad (\alpha >0),   
\end{equation}
 is considered where the evolution of scale factor is considered as a power law in time \cite{Dolgov:2014faa,Ramos:2017cot}.
The rationale behind this model lies in its ability to avoid both the flatness and horizon problems entirely. Furthermore, they are free from the fine-tuning issue \cite{Dolgov1982,Dolgov1997,Ford1987}.

Then the Hubble parameter and the Ricci scalar curvature are written as
\begin{equation}
\label{RicciScalar}
H(t) = {\alpha \over t} ~~~~,~~~~R(t) = 6{\alpha(1-2\alpha) \over t^2}~~.
\end{equation}

From Eq.~(\ref{RicciScalar}) in the limit $ H/m <<1$ (after ignoring higher order terms in $H/m$), the Hubble damping is negligible. So it approximates a harmonic oscillator in expanding spacetime, and with the value of $H(t)$ and $R(t)$ from Eq. (\ref{BG-EOM}), the solution of $\phi(t)$ is written as

\begin{eqnarray}
 & & \phi (t) =  t^{\frac{1}{2}-\frac{3 \alpha }{2}}\left[c_1 J_{\frac{1}{2} (3 \alpha -1) }(m t)+c_2 Y_{\frac{1}{2} (3 \alpha -1)}(m t)\right] ,
   \label{ansatz-BG} \nonumber \\
\end{eqnarray} 
where $J$ and $Y$ are the Bessel functions of the first and second kind, respectively, and $c_1$ and $c_2$ are  dimension full constants.
Substituting:
\[
\quad t = \frac{x}{m} \quad \Rightarrow \quad mt = x \quad \Rightarrow \quad 
t^{\frac{1}{2} - \frac{3\alpha}{2}} = \left( \frac{x}{m} \right)^{\frac{1}{2} - \frac{3\alpha}{2}}
\]
So the full expression becomes:
\[
\quad 
\left( \frac{x}{m} \right)^{\frac{1}{2} - \frac{3\alpha}{2}}
\left[ 
J_{\frac{1}{2}(-1 + 3\alpha)}(x)\, c_1 +
Y_{\frac{1}{2}(-1 + 3\alpha)}(x)\, c_2 
\right]
\]

For large $x$ (axion mass $m$ is very large at the early epoch), the asymptotic forms:
\[
J_\nu(x) \sim \sqrt{\frac{2}{\pi x}} \cos\left( x - \frac{\nu \pi}{2} - \frac{\pi}{4} \right), \quad x \gg 1
\]
\[
Y_\nu(x) \sim \sqrt{\frac{2}{\pi x}} \sin\left( x - \frac{\nu \pi}{2} - \frac{\pi}{4} \right), \quad x \gg 1
\]

\[
\text{where} \quad \nu = \frac{1}{2}(-1 + 3\alpha)
\]

From the Standard cosmology, cold dark matter CDM behaves as

\begin{equation}
\rho_{\scriptsize \text{CDM}} \approx \rho_{\scriptsize \text{CDM},0}\, a(t)^{-3}, 
\end{equation}
So for axion as dark matter, one can write as 
\begin{equation}
\label{rhofr}
\rho_a(t) \approx \rho_{a,0} \left({t \over t_0}\right)^{-3\alpha}~, \quad  p_a(t) = 0.
\end{equation}

This expression represents the energy density $\rho_a(t) $ of  axion that oscillates, with $\rho_{a,0}$ as ${m (c_1^2+c_2^2) \over \pi \, }$, which is a constant. The background medium evolves exactly in the same way as a pressureless ideal fluid; here is CDM (i.e. $\omega=0$)\cite{Turner-1983} and hence $p_a(t)$ is zero.

Hence, for the case, when matter has filled the whole universe and after neglecting of radiation part in comparison to the matter. One can write,
\begin{eqnarray}
\label{matterrho}
\rho_m(t) = \rho_{m,0}\, a(t)^{-3}~, \quad  p_m = 0.
\end{eqnarray} 
where $\rho_m(t) $ is the baryonic matter density, $\rho_{m,0}$ is a constant. Both dark matter and baryonic matter behave the same way. Thus, the total matter density is written as
\begin{eqnarray}
\label{rhofr}
\rho(t) = \rho_a(t) + \rho_m(t) ,   \,\,\,\,\,\rho(t) = \rho_0 \; a(t)^{-3}
\end{eqnarray}

Using Eqs. (\ref{alp}),(\ref{RicciScalar}) and (\ref{rhofr}) into  any of the   Eqs.~(\ref{RAY}-\ref{TR}) and  and comparing the powers of $t$. One can obtains,
\begin{equation}
\label{relation}
\alpha = {2 \over 3} ~.
\end{equation}

Using Eq.~(\ref{ansatz-BG}), and substituting $\alpha=2/3$, we can write solution of axion field  $\phi(t) $ as 
\begin{eqnarray}
   & & \phi (t) = {1 \over t}\sqrt{{2 \over \pi  m}} [c_1 \sin (m t)-c_2 \cos (m t)]
   \label{BG-phi}
\end{eqnarray}  

Substituting the value of $a(t)$,\,$\alpha$, $\rho(t)$, $H(t)$, and $R(t)$ in Eq. (\ref{MFE}), considering suppressed curvature $R/\Lambda \rightarrow 0$ in early universe because around
the inflationary epoch, it is exponentially suppressed by $ \propto e^{-2N}$,
where $N$ is the number of e-folds. For $N \gtrsim 60$, curvature effects become negligible. Then after solving for $\rho_{0}$, 

\begin{equation}
\rho_0 =  {M_P^2 (2 -\beta)\over 3  t_0^2}~,
\end{equation}

By putting the value of $ \rho_0$ into the Eq.~(\ref{rhofr}), one obtains the energy density as:
\begin{equation}
\label{density mn}
\rho(t) =  {M_P^2 (2 -\beta)\over 3  t^2}~,
\end{equation}
\noindent
\begin{figure}[h]
	\begin{centering}
	\begin{tabular}{cc}
  \\ \includegraphics[scale=.97]{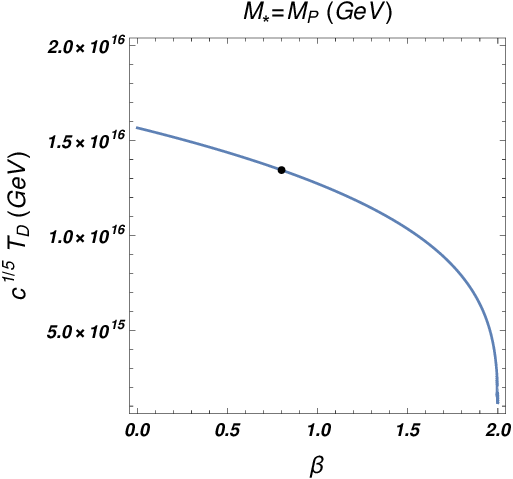}
\end{tabular}
	\end{centering}
\caption{The plot shows the variation of decoupling temperature $T_D$ with $\beta$ at a cut-off scale of $M_* =M_P $ GeV.} \label{ctp1}		
\end{figure}

\begin{figure}[t]
	\begin{centering}
	\begin{tabular}{cc}
	\includegraphics[scale=.96]{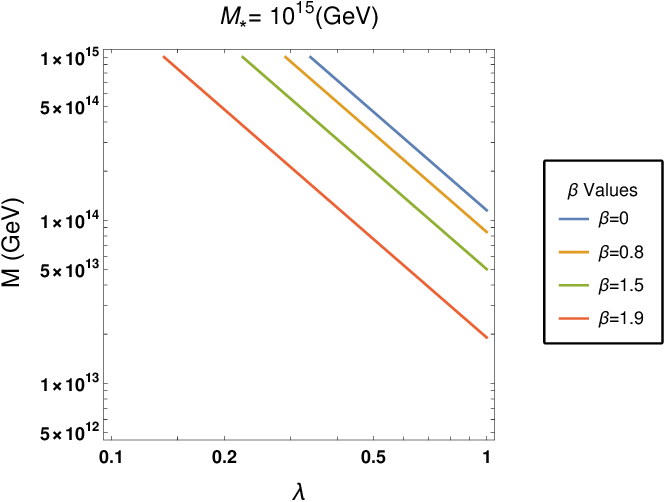} \\
    \includegraphics[scale=.96]{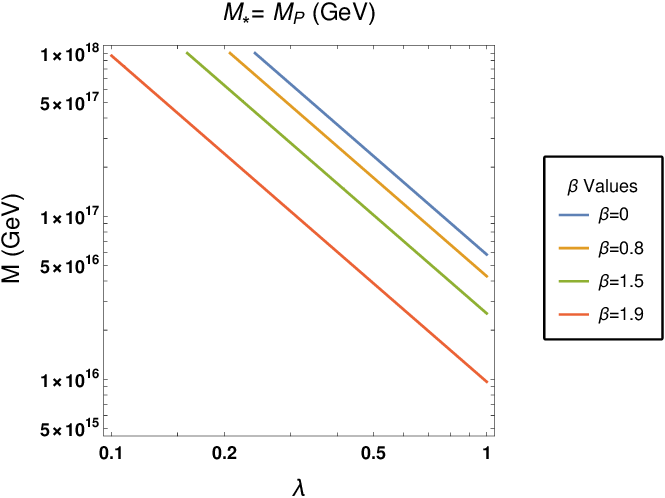}
 \end{tabular}
	\end{centering}
\caption{The plots show the variation between neutrino mass $M$ and Yukawa coupling $\lambda$ with different $\beta$ values at a cut-off scale of $M_*=10^{15}$ GeV ({\it upper}) and $M_*=M_P$ GeV ({\it lower}), respectively.} \label{CDM2}	\end{figure}

Using the relation of matter (baryonic and fermionic) density  $\rho$ at high temperature ($T>>m$) in equilibrium,  one can write as \cite{Baumann2022},
\begin{equation}
\label{deftemp}
\rho = {\pi^2 \over 30}g_* T^4~,
\end{equation}
Comparing with Eq~(\ref{density mn}) temperature is written as,
\begin{eqnarray}
\label{temperature}
T =\left[{ 10 \, M_P^2 (2-\beta) \over g_* \pi^2  t^2}\right]^{1/4}~.
\end{eqnarray}

Using  Eq. (\ref{RicciScalar}) and ~(\ref{temperature}), it follows that
\begin{equation}
\label{rd}
\dot{R} = {8 \over 3 t^3} \approx {2  \sqrt {2} \ T^6 \over 15 \sqrt {5}} \left[{g_* \pi^2 \over  M_P^2 \ (2 -\beta)}\right]^{3/2}~,
\end{equation}
and hence from Eq.(\ref{eta and R}) and (\ref{rd}) the net lepton asymmetry can be written as

\begin{equation}
\label{eta susy}
\eta_l \approx {   \pi   \ g_l \ g_*^{1/2} \over 2 \sqrt{10} } { T_D^5 \over M_*^2} \left[{1  \over  M_P^2 \ (2 -\beta)}\right]^{3/2}_{T = T_D},
\end{equation} 
where $T_D$ is the decoupling temperature, at which the lepton-violating interactions go out of equilibrium. Eq.~(\ref{eta susy}) shows that for leptonic asymmetry to be physical, there will be an upper bound on $\beta < 2$.
The best fit value of $\beta$, when the axion is taken as dark matter,  is $0.8$ \cite{Odintsov:2023nqe}. Interestingly, this is within the upper bound on $\beta$ which we have obtained, so far we have calculated the leptonic asymmetry $\eta_l$. 

In the presence of sphaleron transition, leptonic asymmetry will lead to the creation of baryonic asymmetry in the universe \cite{saphos}. As the transition rate is Boltzmann suppressed and proportional to $e^{\frac{-E_{sph}}{T}}$, where $E_{sph}$ is the barrier height (related to up and down type Higgs vacuum expectation values), the $B+L$ violating sphaleron transition becomes very small near the universe's first-order electroweak phase transition. But $B-L$ is still conserved, because of this, the leptonic asymmetry $\eta_l$ will be converted to baryonic asymmetry $\eta$ \cite{saphos,Harvey1990,Kuzmin1985,Farrar:1993sp,fukugita1986baryogenesis} as : 
\begin{equation}
    \label{c}
    \eta = c \, \eta_l,
\end{equation}
in which $c$ depends on the model in which lepton number-violating interactions are present. The value of the constant $c$ will be discussed separately in the context of specific models having lepton number $L$-violating interactions. In Fig.~\ref{ctp1}, we have shown the variation of decoupling temperature with parameter $\beta$ to get the required baryonic asymmetry. The qualitative feature of the variation of $c^{1/5} T_D$ with $\beta$ can be found in Fig.~\ref{ctp1}. For the higher value of $\beta$, the decoupling temperature $T_D$ decreases. The black dot represents the best fit value of $\beta=0.8$, corresponding to $T_D$ being around $1.4 \times 10^{16}$ GeV.

\subsection{Decoupling Temperature $T_D$ for non SUSY model}

Here, we consider the lepton number violating decay of the heavy right-handed neutrino $N$ to the Higgs field $\phi$ and light active neutrino $\nu_{L\alpha}$ with Yukawa coupling for these fields as $\lambda$. The decay width $\Gamma_N$ for $N$ is 

\begin{eqnarray}\label{neuy}
\Gamma_N = \sum_\alpha \Gamma_{\alpha\alpha} = \sum_\alpha \Gamma(N \rightarrow \phi \,\nu_{L\alpha}) \simeq {{{\lambda^2 M} }\over{8 \pi}},
\end{eqnarray}
and the Hubble parameter $H$ is related to $T$ as follows:
\begin{eqnarray}\label{32}
H = 1.66 g_*^{1/2} \left( {{T^2}\over{M_P}} \right) \Bigg|_{T = M}.  \end{eqnarray}
The $L$ violating decay is decoupled due to the expansion of the universe, for which 

\begin{equation}\label{33}
    \Gamma_N \lesssim H .
\end{equation}
Using Eq.~(\ref{32}) in Eq.~(\ref{33}) and considering approximate equality, one obtains decoupling temperature as

\begin{equation}\label{td1}
  T_D \approx  \lambda  \left[{{M_P M }\over {13.28 \,  \pi  \, g_*^{1/2}}}\right]^{1/2}.
  \end{equation}

In the standard model, the resultant leptonic asymmetry is converted to baryonic asymmetry through sphaleron transition as in \cite{Kuzmin1985,fukugita1986baryogenesis}, for which
\begin{equation}
\eta_b = {1 \over 2} \eta_{l}.
\end{equation}

Here $\eta_b$ is baryonic asymmetry, in this case, the $c$ in Eq.~(\ref{c}) is $1/2$. Using $T_D$ from Eq.~(\ref{td1}), we can rewrite Eq.~(\ref{eta susy}) as,
\begin{equation}
\label{eta nsusy}
\eta_b \approx {2.4 \times 10^{-4} \, g_b  \over   g_*^{3/4} \pi^{3/2} } {\lambda^5 \,   M^{5/2} \over  M_*^2 \; M_P^{1/2}} \left[{1  \over  (2-\beta)}\right]^{3/2}~,
\end{equation} 

In Fig.~\ref{CDM2}, using Eq.~(\ref{eta nsusy}) so that observed baryonic asymmetry could be obtained, we have shown the variation of the heavy right-handed neutrino mass $M$ with the Yukawa coupling $\lambda$ for different values of $\beta$. In the upper plot, $M_* = 10^{15}$ GeV, and in the lower plot, it is $M_P$.
On comparing two plots, one can see that the mass of the right-handed neutrino is relatively higher for the required baryonic asymmetry when the $M_*$ value is higher and with the higher values of $\beta$, both $M$ and $\lambda$ values become lower. However, for getting baryonic asymmetry, the lower value of the right-handed neutrino mass in the upper plot in Fig.~\ref{CDM2} for $M_*$ is about $\sim 10^{13}$ GeV with a Yukawa coupling of the order of $\sim 1$. If this Yukawa coupling is associated with the lightest neutrino, then there is no problem in the Type I seesaw mechanism. But to get other active light neutrino masses, the other two right-handed neutrino masses are required to be less than $M$ to satisfy mass square differences which follow from neutrino oscillation data~\cite{deSalas:2020pgw}. 

\subsection{Decoupling Temperature $T_D$ for SUSY model }

For SUSY, the resultant leptonic asymmetry is converted through the sphaleron transition as \cite{saphos}
\begin{equation}\label{conv}
\eta_b = {{n_{B}-n_{\overline{B}}}\over{s}} = \left( {{8 N_f + 4 N_H} \over { 22 N_f + 13 N_H} }\right) \eta_l
\end{equation}
where $N_H$ is the number of Higgs doublets and $N_f$ is the number of lepton generations in the above equation. 
Considering $N_H =2$ and $N_f =3$ for Minimal Supersymmetric Standard Model
(MSSM), the above relation is,
\begin{equation}
\eta_b={{8}\over{23}}\eta_{l},
\end{equation}
from the above equation, one can see that $c$ in Eq.~(\ref{c}) for MSSM is $8/23$.

The mass and the interaction terms for sneutrinos $\widetilde{N}_+$ and $\widetilde{N}_-$ in the mass eigenstate basis are written as \cite{Adhikari:2015ysa,Khan:2023myt},
\begin{eqnarray}
{\cal L} & = & 
M_{+}^{2}\widetilde{N}_{+}^{*}\widetilde{N}_{+}+M_{-}^{2}\widetilde{N}_{-}^{*}\widetilde{N}_{-}\nonumber \\
 &  & +\frac{1}{\sqrt{2}}\left\{ \widetilde{N}_{+}\left[ Y \overline{\widetilde{H}_{u}^{c}}P_{L}\ell
+\left(A + M Y\right)\widetilde{\ell}H_{u}\right]\right.\nonumber \\
 &  & \left.+i \widetilde{N}_{-}\left[ Y\overline{\widetilde{H}_{u}^{c}}P_{L}\ell
+\left(A- M Y\right)\widetilde{\ell}H_{u}\right] \right\} .
\label{eq:lag}
\end{eqnarray}
where $ A$ is the soft SUSY breaking trilinear term and $Y$ is the Yukawa coupling associated with the interaction of the right-handed neutrino superfield with neutral lepton and Higgs superfields.
The masses of two sneutrinos do not differ much, and we have considered those to be almost equal to $M$.

Based on the interactions in Eq.~(\ref{eq:lag}), the sneutrinos could decay to a lepton and Higgsino or a slepton and a Higgs. The decay width of sneutrinos is given by
\begin{equation}
\label{gamma}
\Gamma  \approx { { M^2 Y^2 + [A + M Y]^2} \over {8  \pi M}}~.
\end{equation}
For decoupling of lepton-violating interactions during the expansion of the universe, like that in the earlier subsection, we consider $ \Gamma \lesssim H$.
Considering approximately equality in this, we obtain $T_D$ as
\begin{equation}
\label{td}
T_D \approx {{M_P}^{1 / 2} \over 1.66 {g_*}^{1 / 4}} \left[{[ A+M Y] ^2+M^2 Y^2} \over  {2 \pi  M}\right]^{1/2}~.
\end{equation}
Using  Eq.~(\ref{td}) we rewrite Eq.~(\ref{eta susy}) as 

\begin{equation}
\label{fult}
\eta_b \approx { 1.2 \times 10^{-2}   \over  {g_*}^{3 / 4}} { \pi   \ g_l \over M_*^2 M_P^{1/2}}  \left[{[ A+M Y] ^2+M^2 Y^2} \over {2 \pi  M \, (2 -\beta)^{3/5} }\right]^{5/2}~,
\end{equation} 

\begin{figure}[h]
	\begin{centering}
	\begin{tabular}{cc}
	\includegraphics[scale=.96]{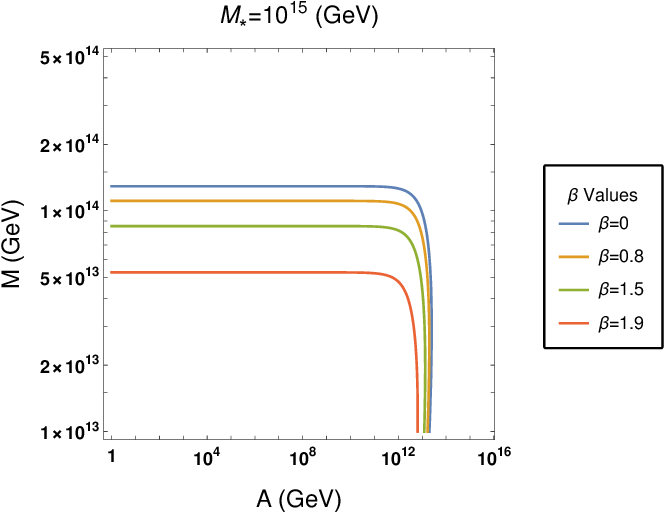} \\
	\includegraphics[scale=.96]{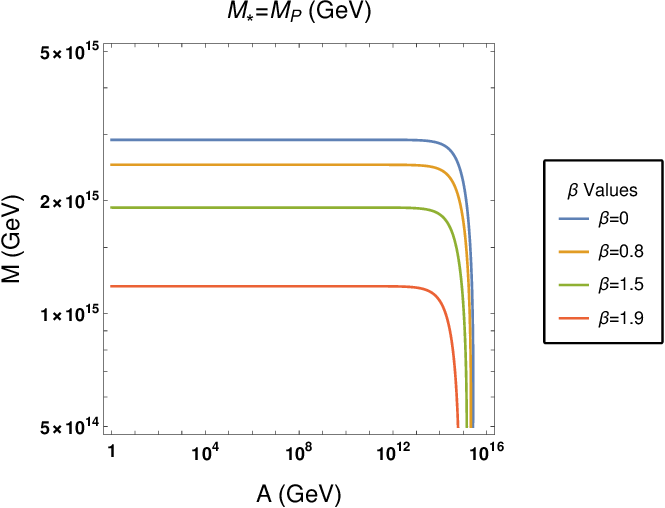}
 \end{tabular}
	\end{centering}
\caption{The plots show the variation between sneutrino mass $M$ and trilinear SUSY breaking parameter $A$  with different $\beta$ values at a cut-off scale of $M_*=10^{15}$ GeV ({\it upper}) and $M_*=M_P$ GeV ({\it lower}), respectively.} \label{cntall}		
\end{figure}

Using the Type I seesaw mechanism in MSSM for obtaining light neutrino mass, one can write 
\begin{equation}\label{neut}
\sum_\alpha {Y_{\alpha}^{2}v_u^{2}\left(1+\rm cot^2 \theta \right) \over M} \approx  
\sum_{i}m_{\nu_{i}} \lesssim 1.18\times 10^{-10} \,{\rm GeV},
\end{equation}
where $v_u^2+v_d^2=\langle H_{u} \rangle $ $\simeq (174 \; \rm GeV)^2$, where $v_u,v_d$  are up-type and down-type Higgs vacuum expectation values, $M$, which is the lightest among the heavy RHN masses, and  $ \rm cot \theta$$ \equiv v_d/v_u $. Considering cot$\theta$ small as compared to $1$, we have neglected this in the later analysis.

In Fig.~\ref{cntall}, we have used Eq.~(\ref{fult}) and the constraint shown in~(\ref{neut}) and have obtained the plot for sneutrino mass, which is $M$, versus the soft-breaking trilinear parameter $A$ for different values of $\beta$ so that observed baryonic asymmetry could be obtained. The variation of the Yukawa coupling $Y$ is not shown in the plot. However, because of the variation of this coupling subject to the constraint in Eq.~(\ref{neut}), it is found that for a wide range of values of the $A$ parameter from $1$ to $10^5$ GeV for almost the same sneutrino mass, it is possible to obtain the required baryonic asymmetry. This is because $M Y >> A$ on the horizontal part of both the upper and lower plots in Fig.~\ref{cntall}. In this case, like the earlier model, one is required to consider the other two right-handed neutrinos to be lighter than $M$.  On the other hand, for higher values of $A$, this is not necessarily true. For higher values of $\beta$, the sneutrino mass is found to be lower in both upper and lower plots in Fig.~\ref{cntall}. 

In the context of the supersymmetric model considered in subsection B, there could be a gravitino problem~\cite{Davoudiasl:2004gf} as the decoupling temperature is high and is above $\sim 10^{10}$ GeV as shown in Fig.~\ref{ctp1}. The decay of a heavy gravitino could affect the result of BBN, or as the lightest supersymmetric particle, it could overclose the universe. For masses of gravitinos above $\sim 100$ TeV, gravitinos are expected to decay before the start of BBN. However, if the lightest supersymmetric particle (which is formed due to gravitino decay) is lighter than $\sim 100$ GeV, then the gravitino problem may be avoided. For a very light mass of gravitino below the keV range, no gravitino problem is expected.

Above $T_D$, any $\Delta L$ produced by the gravitational interaction is washed out due to decays (like 
those mentioned in Eq.~(\ref{neuy}) or just above Eq. (\ref{gamma})) and their inverse decays together with sphaleron induced process in equilibrium. Below $T_D$ due to the out-of-equilibrium scenario, washout will be suppressed. $B+L$ violating sphaleron induced electroweak process will not wash out any $\Delta L$ which will be converted to baryonic asymmetry due to $B-L$ conservation in the electroweak process.

\section{Discussion and conclusion}
\label{dis}
In this paper, we have shown the generation of leptonic asymmetry in exponential $f(R)$ gravity. The lepton asymmetry comes from the fact that the decay of a heavy right-handed neutrino or sneutrino violates the lepton number. Through gravitational interaction with leptonic current in Eq.~(\ref{int}), the asymmetry is generated in thermal equilibrium and is related to the time derivative of the scalar curvature. In our work, this derivative depends on the exponential $f(R)$ gravity term and the $\rho_a(t)$ due to cold dark matter axion.

The leptonic asymmetry, as shown in Eq.~(\ref{eta susy}), shows that there should be an upper bound on the parameter $\beta$ to be less than $2$. Interestingly, the best fit value of $\beta$ for late time acceleration, which is found to be $0.8$ \cite{Odintsov:2023nqe}, is within this upper bound. In this sense, the gravitational baryogenesis could be reconciled with exponential $f(R)$ gravity. 

The leptonic asymmetry in Eq.~(\ref{eta susy}) also indicates that for obtaining the observed baryonic asymmetry, the decoupling temperature is required to be high. This feature is there in the original gravitational baryogenesis work \cite{Davoudiasl:2004gf}, and inclusion of the $f(R)$ gravity term and the cold dark matter does not change that feature as expected. However, there is a small variation of the decoupling temperature $T_D$ with the variation of the $\beta$ parameter. In the context of two models with lepton-number-violating interactions, the decoupling temperature has been written in terms of model parameters. As the decoupling temperature varies with the $\beta$, using that relationship, we have shown how model parameters related to lepton number violating interactions also could vary with the $\beta$, although such variations, as shown in Figs.~\ref{CDM2} and~\ref{cntall}, are found to be small.

In principle, one may consider either leptonic current or baryonic current in the interaction in Eq.~(\ref{int}). Based on that, either leptonic asymmetry or baryonic asymmetry will be produced through Eq.~(\ref{eta and R}). However, such asymmetry will be proportional to $T_D^5$ as shown in Eq.~(\ref{eta susy}). Correspondingly, the decoupling temperature $T_D$ depends on either $L$ or $B$ violating couplings. But for $B$ violating coupling, there is a stringent constraint that comes from the stability of the universe due to possible proton decay. For  $L$ violating coupling, there is no such stringent constraint associated with the stability of the universe, and appropriate asymmetry through Eq.~(\ref{eta susy}) could be easily generated. So we have preferred to study the gravitational interaction with leptonic current in Eq.~(\ref{int}).
 
 \noindent 
\section*{Acknowledgements}\label{sec:acknowl}
SK thanks the Council of Scientific and Industrial Research (CSIR), India, for financial support through
Senior Research Fellowship (Grant No. 09/466(0209)/2018-EMR-I). SK and RA sincerely thank the anonymous referees for their valuable suggestions, 
which have greatly improved the quality and clarity of the manuscript.

\bibliography{fr}
\bibliographystyle{apsrev4-1}
\end{document}